\documentclass{appolb}
\usepackage{epsfig}
\usepackage{amsmath}
\usepackage{amssymb}
\usepackage{slashed}
\usepackage{color}

\definecolor{refcol}{RGB}{178,34,34}
\usepackage[colorlinks,linkcolor=refcol,citecolor=refcol,urlcolor=refcol]{hyperref}

\newcommand{\um}{\ensuremath{{U^{\text{mes}}}}}

\graphicspath{
{./} 
 }

\begin{document}
\title{Chiral Thermodynamics in a finite box
\thanks{Presented by B.-J. Schaefer at "Critical Point and Onset of Deconfinement"
(CPOD2016), Wroc\l{}aw, Poland; May 30$^{\text{th}}$ - June
4$^{\text{th}}$, 2016.}
}
\author{Ana Juri\v{c}i\'{c}$^{1,2}$ and  Bernd-Jochen Schaefer$^{2}$
\address{$^{1}$Institut f\"ur Physik, Universit\"at Graz, Universit\"atsplatz 5, 8010
Graz, Austria\\
$^{2}$Institut f\"ur Theoretische
  Physik, Justus-Liebig-Universit\"at Gie\ss en, Heinrich-Buff-Ring 16, 35392 Gie\ss
  en, Germany}
%
}
\maketitle
\begin{abstract}
  Finite-volume modifications of the two-flavor chiral phase diagram
  are investigated within an effective quark-meson model in various
  mean-field approximations.  The role of vacuum fluctuations and
  boundary conditions, their influence on higher cumulants and
  signatures of a possible pseudo-critical endpoint are amplified with
  smaller volumes.
\end{abstract}

\section{Introduction}

At sufficiently high temperature and densities QCD predicts a phase
transition from a hadronic phase to a deconfined quark-gluon
plasma. Obviously, a plasma created in a laboratory experiment is
restricted to a finite (small) volume surrounded by a cold
exterior. These volumes are not of a fixed size and one has to deal
with ensembles of differently shaped and sized volumes depending on
e.g. the centrality of a heavy-ion collision. Thermodynamic quantities
directly inferred from such data should not correspond to theoretical
predictions obtained in the thermodynamic limit where non-negligible
finite-size corrections are ignored.  Furthermore, any system of
finite size cannot exhibit a real sharp phase transition and
criticality cannot be observed in the usual sense. This regards, in
particular, the singularity connected with the existence of a
second-order critical endpoint (CEP) in the QCD phase diagram
\cite{Braun:2005gy}. All critical quantities become pseudocritical
with certain ambiguities in their definitions.

The scope of this talk is to address finite-size effects within a
well-defined low-energy effective model for the two-flavor chiral
phase transition \cite{Palhares:2009tf}.  We employ different
approximations of the grand potential to study the influence of
quantum and thermal fluctuations on the phase transition in different
volumes. Even though the actual obtained numbers in such an effective
model analysis should not be overrated the generic statements relative
to the thermodynamic limit are of relevance which turn out to be of
significance.
We adopt the two-flavor quark-meson model 
\begin{equation}
  \label{eq:8}
  \mathcal{L} = \bar{q} \left( i \slashed{\partial} - g (\sigma 
    + i \vec{\tau} \cdot \vec{\pi} \gamma_5) +\mu \gamma^0 \right) q 
  + \frac{1}{2}(\partial_{\mu} \phi)^2 - \um (\phi)
\end{equation}
where the meson fields are conflated in the $O(4)$-symmetric
vector $\phi^T = (\sigma, \vec{\pi})$ with a purely mesonic potential
\begin{equation}
  \label{eq:35}
  \um (\phi) =  \frac{\lambda}{4} (\phi^2 - v^2)^2 - c \sigma\ .
\end{equation}
The four meson fields are coupled to the quark fields $q^T = (u,d)$
via a Yukawa coupling $g$. Chiral symmetry is broken explicitly by a
linear $\sigma$-term in the potential. The partition function of the
model is given by a path integral over all quark and meson fields
which cannot be done analytically. In mean-field approximations the
path integration is approximated by performing only the quark
integration and ignoring the meson integrals. The quark loop yields a
generically divergent vacuum contribution to the resulting
thermodynamic grand potential. By choosing a regularization scheme
with a certain ultraviolet cutoff $\Lambda$ we can investigate the
influence of these vacuum fluctuations on the system.  In standard
mean-field approximation (sMFA) the vacuum term to the potential is
simply ignored which corresponds to setting $\Lambda=0$.  Since the
quark-meson model is a renormalizable model one can fully include the
vaccum term by sending $\Lambda \to \infty$ which we will denote as
renormalized mean-field approximation (rMFA) and goes beyond the usual
sMFA. In this way vacuum and thermal fluctuations of the quarks are
taken into account. In contrast, vacuum and thermal fluctuations of
the meson fields are always ignored in such mean-field approximations.

\section{Finite volume effects}

For our investigation of a system in a finite three-dimensional volume
we choose a spatial box of equal length $L$. The temporal direction is
also compactified to finite size $\beta= 1/T$ to incorporate finite
temperature.  The boundary conditions in the temperature direction are
determined by the spin-statistics theorem and provide the usual
thermal Matsubara frequencies for bosons and quarks. In contrast, the
boundary conditions in the spatial directions can be chosen
independently for bosons and fermions. This freedom is utilized in
lattice Monte Carlo simulations where usually periodic boundary
conditions for quarks in spatial direction are employed since they
ensure a fast approach to the infinite volume limit. Moreover, this
enables a further comparison of the influence of the chosen boundary
conditions on the phase structure.  Despite possible generalization to
arbitrary phases only periodic (PBC), periodic without the zero mode
(PBC*) and antiperiodic boundary conditions (ABC) where the zero mode
is absent are considered in the following.

\begin{figure}[htb] 
\includegraphics[scale=0.5] {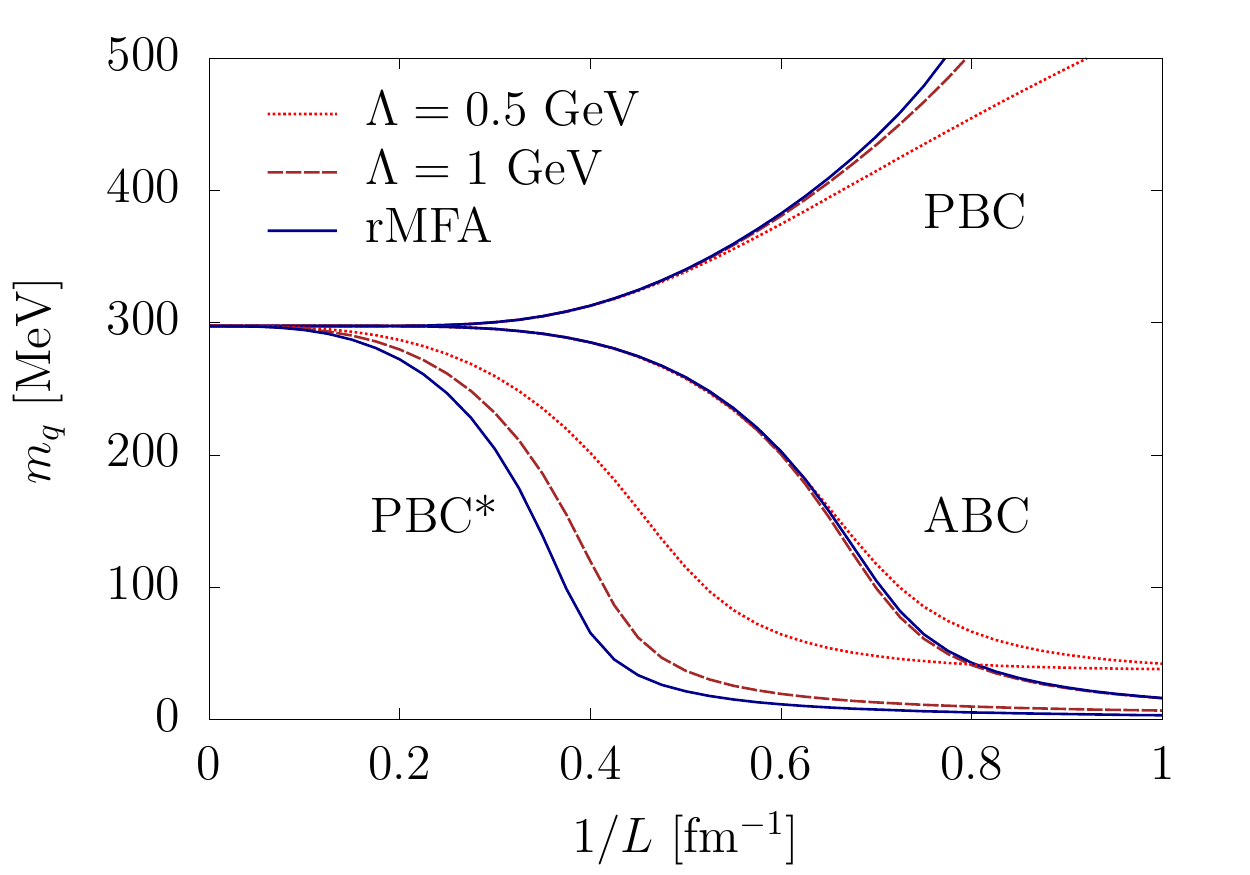}
\includegraphics[scale=0.5] {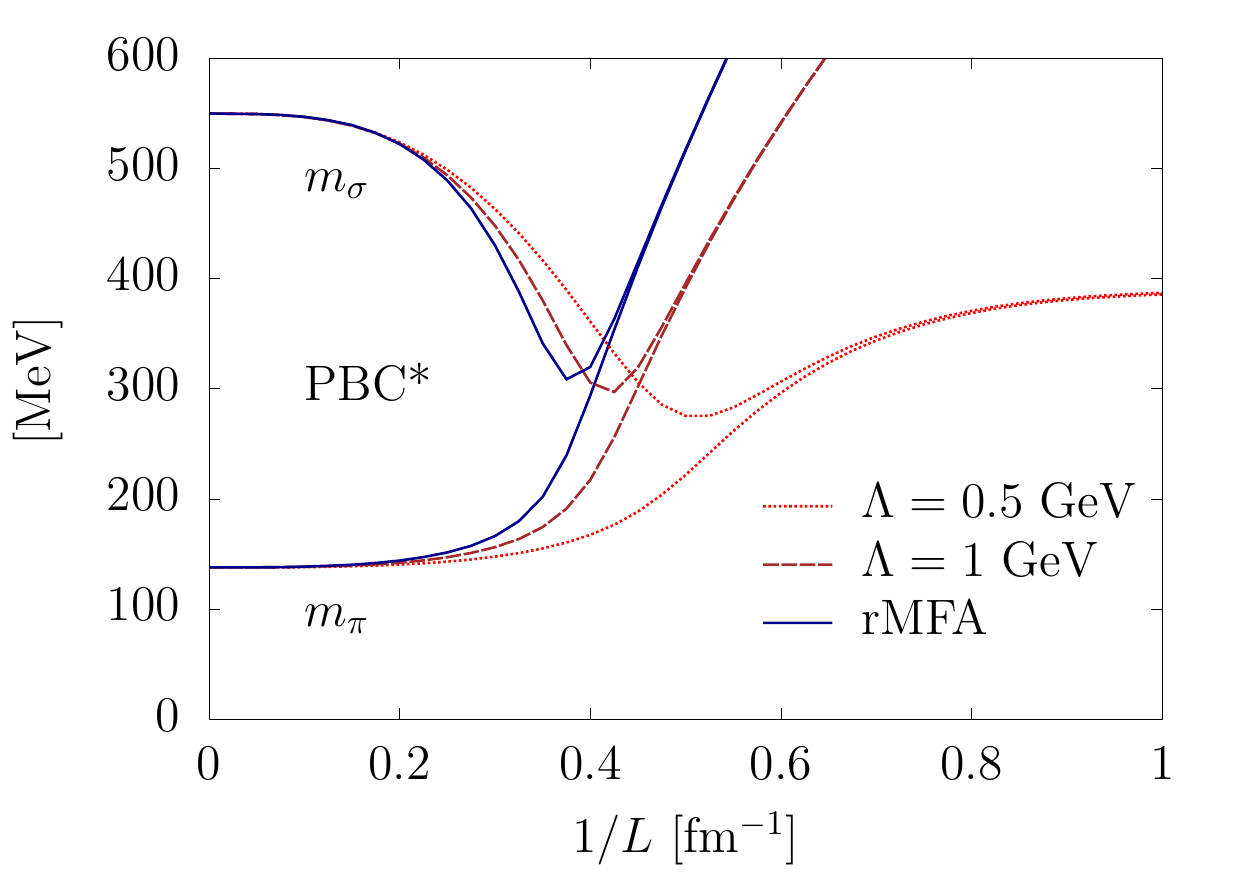}
\caption{Quark (left panel) and meson masses (right panel) as a
  function of the inverse system size for different UV-cutoffs and
  boundary conditions.}
\label{fig:masses}
\end{figure} 

In Fig.~\ref{fig:masses} the quark masses (left panel) and the
curvature meson masses (right panel) are shown as a function of the
inverse size. Without the zero mode all masses display a similar
behavior as at finite temperature which demonstrates the analogy of
the temperature with the inverse size. For decreasing volumes and PBC*
a chiral symmetry restoration "phenomenon" around $1/L \sim 0.4$
fm$^{-1}$ is found which is shifted to smaller volumes for
ABC. Moreover, the vacuum fluctuations affect the $L$-dependence
stronger in comparison to ABC.  This behavior is in contrast to the
unphysical PBC case when the zero mode is included and is illustrated
in the left panel of Fig.~\ref{fig:masses}.

\begin{figure}[htb!]
\includegraphics[scale=0.5]{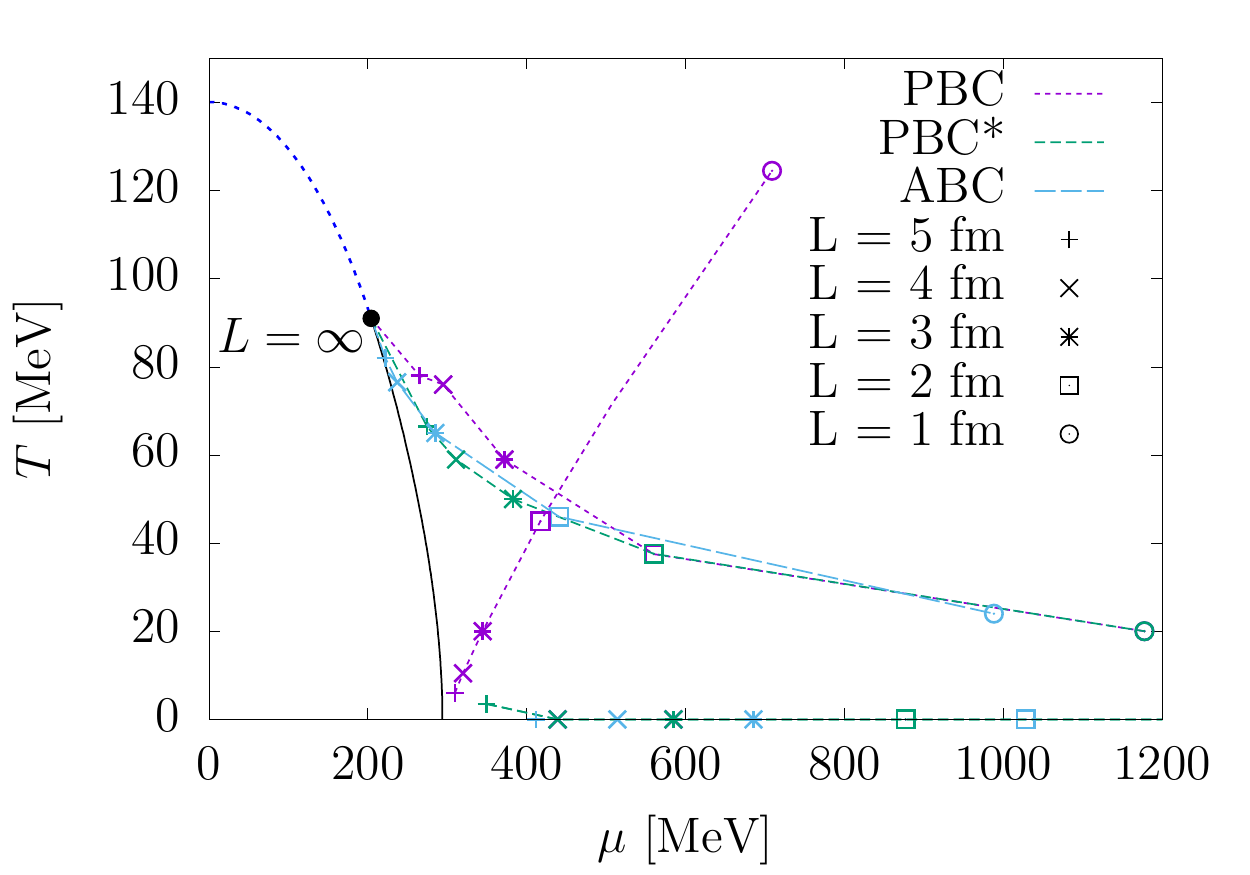}
\includegraphics[scale=0.5]{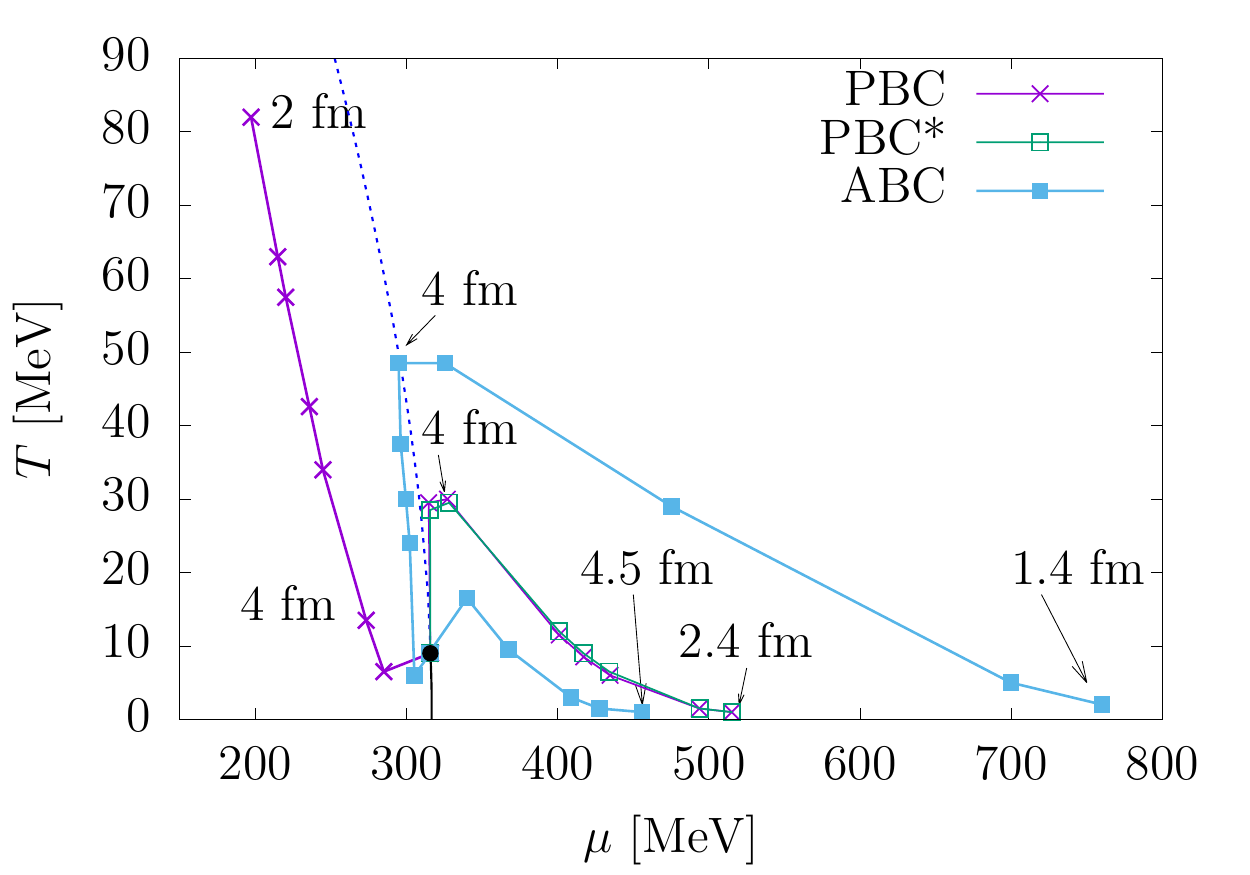}
\caption{Size sensitivity of the pseudocritical point for different
  boundary condition. Left panel without (sMFA) and right panel with
  vacuum fluctuations (rMFA).}
\label{fig:endpoints}
\end{figure}

In Fig.~\ref{fig:endpoints} the size-sensitivity of the pseudocritical
endpoints for three different spatial boundary conditions are shown
and compared to the infinite volume phase diagram obtained in the
thermodynamic limit (solid black line represents the first-order
transition). The left panel represents the shift of the location
without the vacuum fluctuations (sMFA) while these fluctuations are
included in the right panel (rMFA). Including the zero mode (PBC)
additional discontinuities in the order parameter emerge and as a
consequence further endpoints of these discontinuity-lines appear. In
addition for small temperatures more discontinuities in the order
parameter are visible which we also include in the figure.  However,
only one endpoint per volume size is connected to the crossover line
which we identify as the physical pseudocritical endpoint.  Excluding
the zero-mode almost no difference between the boundary conditions
(ABC) and (PBC*) are found. Both boundary conditions reproduce the
infinite-volume limit already for sizes $L \geq 10$ fm and for smaller
volumes the location is pushed to higher chemical potentials. This
changes drastically when vacuum fluctuations are taken into
account. In rMFA the findings for both boundary conditions coincide
for $L \geq 10$ fm but start to deviate by finite-size corrections. At
first the pseudocritical endpoint moves upwards to higher temperatures
and then turns down and vanishes at larger chemical potentials from
the phase diagram. If this scenario is a generic one this would be bad
news for the current experiments which may not be able to probe this
regime. It is also interesting to see finite-size modifications in the
crossover temperature at vanishing chemical potential if vacuum
fluctuations are considered or not. In sMFA the crossover temperature
at $\mu=0$ increases with decreasing system sizes for all used
boundary condition strongest for the one without the zero mode
(PBC*). In rMFA no significant change is seen for the PBC and ABC in
contrast to the one when the zero mode is excluded. For this case the
crossover temperature decreases about $20\%$ for smaller systems and
turns back for $L \lesssim 2$ fm towards larger values again.

These findings seem to agree with results obtained with the functional
renormalization group in a similar quark-meson model truncation where
additionally mesonic fluctuations are taken into account. Depending on
the pion mass the crossover curvature decreases for decreasing volume
sizes for PBC. However, below a certain length scale
$m_{\pi} L \lesssim 2$ the curvature turns back and increases again
for smaller volumes and even exceeds its infinite-volume value which
could be traced back to the spatial zero mode. On the other hand, for
ABC the curvature is a monotonically decreasing function of the volume
size.  This behavior, in turn, has the consequence that the critical
point is pushed towards larger values of the (isospin symmetric) quark
chemical potential and smaller temperatures for PBC and sizes in
between $2 \lesssim m_{\pi}L < \infty$~\cite{Braun:2011iz}.

\begin{figure}[htb]
\includegraphics[scale=0.5]{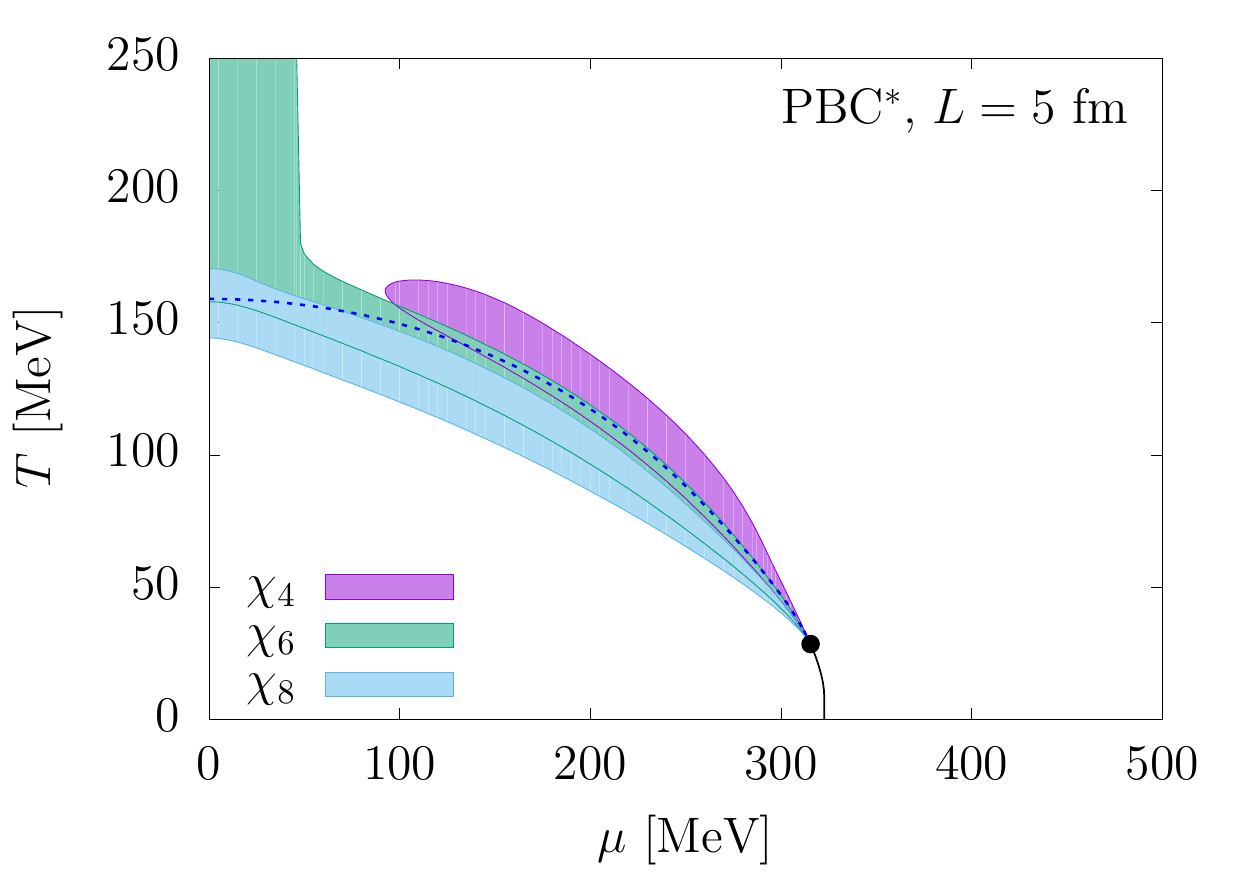}
\includegraphics[scale=0.5]{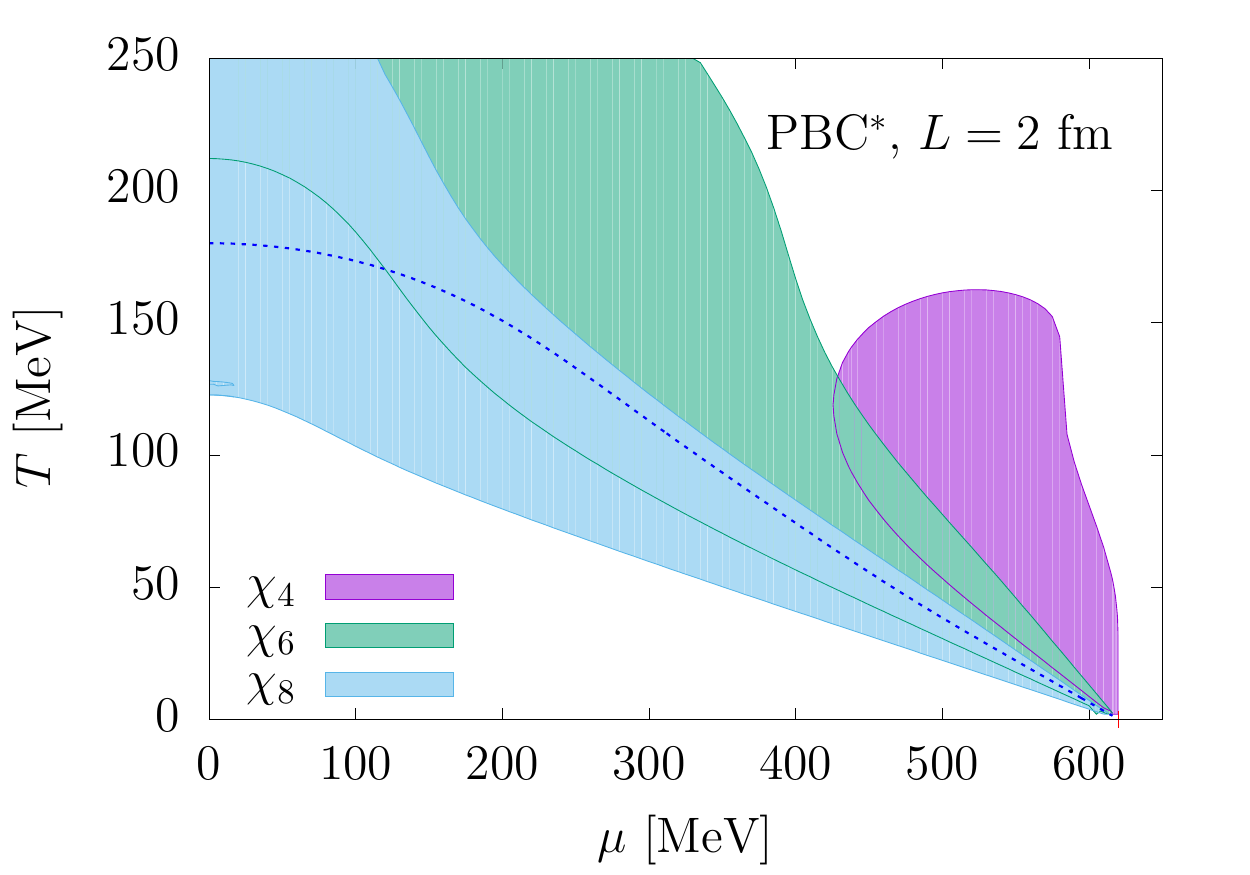}
\caption{First three even cumulants $\chi_4$, $\chi_6$ and $\chi_8$ in
  the finite-size phase diagram excluding the zero mode in rMFA (left
  $L=5$ fm; right $L=2$ fm).}
\label{fig:cumulants}
\end{figure}

In the vicinity of the CEP the scaling behavior can be applied to
infer its presence by employing event-by-event analyses in heavy-ion
experiments. Higher cumulants of particle multiplicities fluctuations
increase with higher powers of the correlation length.  Finite-size
corrections smooth out such singularities and only a moderate increase
is visible. This in particular involves the region where there is a
sign change in higher cumulants. In Fig.~\ref{fig:cumulants} the
negative region of the first three even cumulants are shown in the
finite-size phase diagram obtained in rMFA. Around $L \sim 2.4$ fm the
CEP is gone and the negative region of all cumulants are drastically
washed out. Without the vacuum fluctuations the smoothening is more
moderate and the negative regions focus around the crossover
transition.

\noindent 
\emph{Acknowledgements} This work has been supported by the FWF grant
P24780-N27 and the Helmholtz International Center for FAIR within the
LOEWE program of the State of Hesse.

\bibliographystyle{bibstyle}

\end{document}